\documentclass[aps,prd,twocolumn,nofootinbib]{revtex4-1}
\usepackage{epsfig}
\usepackage[colorlinks,linkcolor=blue,anchorcolor=blue,citecolor=blue,urlcolor=blue,breaklinks=true]{hyperref}
\usepackage{graphics}
\usepackage{slashed}
\usepackage{color}
\usepackage{amsmath}
\usepackage{bm}
\usepackage{setspace}
\usepackage{caption}
\usepackage{multirow}
\usepackage{subfigure}

\begin{document}
\author{Cheng-Ming Li$^{1,6}$}\email{licm.phys@gmail.com}
\author{Yan Yan$^{2}$}\email{2919ywhhxh@163.com}
\author{Jin-Jun Geng$^{3}$}\email{gengjinjun@nju.edu.cn}
\author{Yong-Feng Huang$^{3}$}\email{hyf@nju.edu.cn}
\author{Hong-Shi Zong$^{1,4,5,6}$}\email{zonghs@nju.edu.cn}
\address{$^{1}$ Department of Physics, Nanjing University, Nanjing 210093, China}
\address{$^{2}$ School of mathematics and physics, Changzhou University, Changzhou, Jiangsu 213164, China}
\address{$^{3}$ School of Astronomy and Space Science, Nanjing University, Nanjing 210023, China}
\address{$^{4}$ Joint Center for Particle, Nuclear Physics and Cosmology, Nanjing 210093, China}
\address{$^{5}$ State Key Laboratory of Theoretical Physics, Institute of Theoretical Physics, CAS, Beijing, 100190, China}
\address{$^{6}$ Nanjing Proton Source Research and Design Center, Nanjing 210093, China}
\title{Constraints on the hybrid equation of state with a crossover hadron-quark phase transition in the light of GW170817}

\begin{abstract}
In this paper, we use the recent updated source properties of GW170817 to constrain the hybrid equation of state (EOS) constructed by a three-window modeling between the hadronic EOS and quark EOS. Specifically, the hadronic EOS is described by NL3$\omega\rho$ model whose corresponding pure neutron star (NS) is already excluded by the constraint of tidal deformability (TD) from GW170817, and the quark EOS is calculated with 2+1 flavors Nambu-Jona-Lasinio (NJL) model. We also consider other four constraints on the hybrid EOS. As a result, we find the parameter set ($B^{\frac{1}{4}}, \tilde{\mu}, \Gamma$) can be well constrained, indicating the possible existence of the hybrid star (HS) with a crossover inside. The type of the two stars in the binary system for nine representative hybrid EOSs is shown in this paper too. Furthermore, the HSs restricted by five constraints do not suggest a pure quark core but a mixed-phase in center.

\bigskip

\bigskip


\end{abstract}


\maketitle

\section{INTRODUCTION}
The simultaneous direct detection of the gravitational wave (GW) and its electromagnetic counterpart by LIGO-VIRGO collaboration~\cite{PhysRevLett.119.161101} and $\sim$70 astronomical detectors~\cite{2041-8205-848-2-L12} opens a new era of multi-messenger astronomy. All these observations indicate that the event GW170817 is related to a binary neutron star(BNS) merger. Many works have followed up after that. In paper~\cite{2041-8205-848-2-L13}, the central engine of the short gamma ray burst (GRB) has been studied; while in paper~\cite{2041-8205-850-2-L39}, the study of heavy elements as well as their abundance in the universe has been done. In addition to that, the internal structure of NSs has also been studied with thorough analysis of the new data~\cite{PhysRevLett.120.172703,PhysRevLett.120.172702,2041-8205-850-2-L19,2041-8205-850-2-L34,PhysRevD.97.084038,PhysRevD.96.123012,PhysRevD.97.083015,PhysRevD.97.021501,2041-8205-852-2-L29,2041-8205-852-2-L25,0004-637X-857-1-12,2018ApJ...862...98Z,2018ApJ...860...57A,Ma:2018jze}, but definitive answers are still difficult to find.

It is believed that with more observations of GW events in the future, a better understanding and constraint on the EOS can be achieved, thus considerably promoting research on dense nuclear matter physics~\cite{PhysRevLett.120.031102}. In fact, during the inspiral phase, a star can exert a static tidal field on its companion in the binary, and the quadrupolar response of the field is relevant to the EOS-dependent TD parameter. In papers~\cite{PhysRevD.77.021502,0004-637X-677-2-1216,PhysRevD.81.123016,PhysRevLett.120.261103}, the authors demonstrate the connection between this parameter and the inspiral signal of GW. From the observation data of GW170817, the LIGO-VIRGO collaboration provided a constraint on the dimensionless TD for 1.4 $M_{\odot}$ as $\Lambda(1.4M_{\odot})\leq800$~\cite{PhysRevLett.119.161101}. The upper limit is revised to be 900 for a low-spin prior in the recent paper~\cite{Abbott:2018wiz}. The restriction considerably influences the study of pure hadronic NSs~\cite{PhysRevLett.120.172702}, quark stars~\cite{PhysRevD.97.083015}, and HSs~\cite{PhysRevD.97.084038,0004-637X-857-1-12}. It is noteworthy that for the study of HSs, different aspects and approaches to hadron-quark phase transition will lead to different results. For example, in Ref.~\cite{PhysRevD.97.084038}, a first-order phase transition is considered with the parametrization approach; and in Ref.~\cite{0004-637X-857-1-12} a smooth phase transition with the Gibbs construction is adopted.

Different from the Gibbs construction, the three-window interpolating approach corresponds to a crossover hadron-quark phase transition and the EOS of which can be differentiated to infinite order during the transition region. In addition to that, this interpolating approach is feasible especially when we demand a mall radii of NSs, i.e. $R\lesssim13$ km, or the EOS to be soft at low density but stiff at high density~\cite{KOJO2016821}. Considering the possibility of HSs with a crossover between hadronic matter and quark matter inside~\cite{Masuda01072013,0004-637X-764-1-12,PhysRevD.92.054012,PhysRevC.93.035807,KOJO2016821,PhysRevD.95.056018,PhysRevD.97.103013}, it is reasonable to evaluate the influence of TD parameter on stars of this type. Thus, in this paper, we will investigate the constraint on HSs constructed by the three-window interpolating approach~\cite{Masuda01072013,0004-637X-764-1-12} to connect the quark phase and hadronic phase, which is described by 2+1 flavors NJL model~\cite{RevModPhys.64.649,Buballa2005205,BUBALLA200436,KLAHN2007170,PhysRevD.94.094001} and relativistic mean field (RMF) NL3$\omega\rho$ model~\cite{PhysRevLett.86.5647,PhysRevC.94.035804}, respectively. It is noteworthy that many studies~\cite{0004-637X-764-1-12,PhysRevD.92.054012,PhysRevC.93.035807,KOJO2016821,PhysRevD.95.056018} with this approach have obtained good results for the mass of HSs, namely, the maximum mass compatible with 2 $M_{\odot}$. However, the choice of the interpolating parameters ($\tilde{\mu}, \Gamma$) seems somewhat arbitrary in relevant studies. With the recent updated source properties of GW170817~\cite{Abbott:2018wiz} as well as other four constraints (the mass constraint from PSR J0348+0432~\cite{Antoniadis1233232}, the studies of hadron-quark transition in Refs.~\cite{PhysRevD.77.114028,0034-4885-74-1-014001} implying that $\mu_{\rm deconfinement}>\mu_{\rm ChiralRestoration}\sim1$ GeV at zero temperature with finite chemical potential, the stability of hybrid EOSs~\cite{PhysRevC.93.035807}, the stability of the heaviest HS) on hybrid EOSs, we try to restrict the parameter space and demonstrate the type of two stars in the binary for nine representative hybrid EOSs.

The article is organized as follows: In Sec.~\ref{one}, we present the EOS of hadronic matter at low densities and calculate the EOS of quark matter at high densities. A link between the two phases via three-window interpolating approach is also introduced. Then the methods of constraining parameters are presented in Sec.~\ref{two}. In Sec.~\ref{three}, we give the result of hybrid EOSs and the restricted parameter space of it. A brief summary and discussion are provided in Sec.~\ref{four}. Finally, detailed derivations and calculations of quark condensate are presented in the Appendix~\ref{five}.

\section{CONSTRUCTION OF THE HYBRID EOS}\label{one}
\subsection{EOS of hadronic matter}
The RMF model NL3$\omega\rho$~\cite{PhysRevLett.86.5647,PhysRevC.94.035804} is very successful in describing the confined hadronic matter in beta-equilibrium. The Lagrangian of it reads
\begin{widetext}
\begin{eqnarray}
\mathcal{L}&=&\sum_{N=p,n}\bar{\psi}_N[\gamma^{\mu}(i\partial_{\mu}-g_{\omega N}\omega_{\mu}-\frac{g_{\rho N}}{2}\tau\cdot\rho_{\mu})-(m_N-g_{\sigma N}\sigma)]\psi_N\,\nonumber\\
&+&\frac{1}{2}\partial_{\mu}\sigma\partial^{\mu}\sigma-\frac{1}{2}m^2_{\sigma}\sigma^2-\frac{1}{4}\Omega^{\mu\nu}\Omega_{\mu\nu}+\frac{1}{2}m^2_{\omega}\omega_{\mu}\omega^{\mu}-\frac{1}{4}\rho^{\mu\nu}\cdot\rho_{\mu\nu}+\frac{1}{2}m^2_{\rho}\rho^{\mu}\cdot\rho_{\mu}\,\nonumber\\
&-&\frac{1}{3}b m_{N}(g_{\sigma N}\sigma)^3-\frac{1}{4}c(g_{\sigma N}\sigma)^4+\Lambda_{\omega}(g^2_{\omega}\omega_{\mu}\omega^{\mu})(g^2_{\rho}\rho_{\mu}\cdot\rho^{\mu}).\,\,\,\,\label{hadroniclagrangian}
\end{eqnarray}
\end{widetext}

Compared with the RMF model NL3, this Lagrangian has one more term, i.e., nonlinear $\omega\rho$ term, resulting in softer dependence of the symmetry energy on density. In addition, the exclusion of a quartic term on $\omega$-meson makes the EOS of NL3$\omega\rho$ model very stiff at large densities. Thus the neutron star constructed by NL3$\omega\rho$ has a very large maximum mass, which is calculated to be about 2.75 solar mass ($M_{\odot}$), well above the 2.01$\pm$0.04 M$_{\odot}$ constraint of PSR J0348+0432~\cite{Antoniadis1233232}. In Ref.~\cite{PhysRevLett.86.5647}, we can see from calculations of microscopic neutron matter that this model is compatible with various critical constraints: theoretical, experimental and astrophysical. The saturation properties of NL3$\omega\rho$ are shown in the following: saturated density $\rho_0=$0.148 fm$^{-3}$, energy per nucleon $E/A=$-16.2 MeV, incompressibility $K=$271.6 MeV, symmetry energy $J=$31.7 MeV, slope of symmetry energy $L=$55.5 MeV.

It is known that the structure of a neutron star can be divided into four parts, that is, the envelope, the outer crust, the inner crust and liquid core as the energy density increases. The envelope of the neutron star with energy density smaller than 10$^6$ g/cm$^3$ possesses a tiny mass (10$^{-10}$ $M_{\odot}$), and its conformation and structure can also be affected by many factors such as strong magnetic field~\cite{1994esa..book.....C} and the accretion of interstellar matter. Therefore, in this paper, we will restrict our calculation to $\epsilon>10^6$ g/cm$^3$. Then to build an EOS for the hadronic matter, in the outer crust where $\rho<3\times10^{-4}$ fm$^{-3}$, we employ the Baym-Pethick-Sutherland (BPS) EOS which describes the nuclear matter in this region quite well; in the inner crust and the core where $\rho>3\times10^{-4}$ fm$^{-3}$, we adopt NL3$\omega\rho$ EOS which characterizes the properties of hadronic matter in this region very well. In the meanwhile, these two EOSs intersect at the density of $3\times10^{-4}$ fm$^{-3}$. As a result, the maximum mass of neutron star calculated by this hadronic EOS is about 2.754 $M_{\odot}$ with a radius $R=13.01$ km, implying a very small mass of the outer crust too. In addition, we do not consider the contribution of hyperons in this paper because the interactions among them are complicated and still unknown.

\subsection{EOS of quark matter}
The Lagrangian of 2+1 flavors NJL model has a general form as
\begin{eqnarray}
\mathcal{L}=&&\bar{\psi}(i{\not\!\partial}-m)\psi+\sum^8_{\rm i=0}G[(\bar{\psi}\lambda_{\rm i}\psi)^2+(\bar{\psi}i\gamma_{5}\lambda_{\rm i}\psi)^2]\nonumber\\
&&-K\,({\rm det}[\bar{\psi}(1+\gamma_{5})\psi]+{\rm det}[\bar{\psi}(1-\gamma_{5})\psi]),\,\,\label{lagrangian}
\end{eqnarray}
here $G$ and $K$ are four-fermion and six-fermion coupling constant, respectively; $\lambda^{\rm i}, {\rm i}=1\rightarrow 8$ is the Gell-Mann matrix and $\lambda^0=\sqrt{\frac{2}{3}}\,I$ ($I$ is the identity matrix). In this model, the quark propagator $S_{\rm i}$ can be expressed as
\begin{equation}\label{propagator}
  S_{\rm i}(p^2) = \frac{1}{\not\!p-M_{\rm i}},
\end{equation}
where the subscript i$=u, d, s$ denotes the flavor of the quark and $M_{\rm i}$ represents the constituent quark mass. Then the gap equation can be derived with the mean field approximation as
\begin{equation}\label{udsqge}
  M_{\rm i} = m_{\rm i}-4G\langle\bar{\psi}\psi\rangle_{\rm i}+2K\langle\bar{\psi}\psi\rangle_{\rm j}\langle\bar{\psi}\psi\rangle_{\rm k}.
\end{equation}
Here $\langle\bar{\psi}\psi\rangle_{\rm i}$ and $m_{\rm i}$ are the quark condensate and current quark mass of flavor i respectively, and (i, j, k) is a permutation of ($u, d, s$). On account of the isospin symmetry between u and d quark in 2+1 flavors NJL model, we can obtain that $M_{\rm u}=M_{\rm d}$, $\langle\bar{\psi}\psi\rangle_{\rm u}=\langle\bar{\psi}\psi\rangle_{\rm d}$ and $m_{\rm u}=m_{\rm d}$. By definition, the quark condensate is
\begin{eqnarray}
  \langle\bar{\psi}\psi\rangle_{\rm i} &=& -\int\frac{{\rm d}^4p}{(2\pi)^4}{\rm Tr}[iS^{\rm i}(p^2)]\nonumber\\
  &=& -N_{\rm c}\int_{-\infty}^{+\infty}\frac{{\rm d}^4p}{(2\pi)^4}\frac{4iM_{\rm i}}{p^2-M_{\rm i}^2}.\,\,\label{qcondensate}
\end{eqnarray}
The trace "Tr" is performed in Dirac and color spaces. To proceed with the following calculation, we will make a Wick rotation from Minkowski space to Euclidean space and introduce the Proper Time Regularization (PTR). After that, a generalization from zero temperature and chemical potential to zero temperature but finite chemical potential will be made. The detailed definition and derivation can be found in the Appendix~\ref{five}. Then the quark condensate becomes
\begin{widetext}
\begin{eqnarray}
   \langle\bar{\psi}\psi\rangle_{\rm i} &=& \left\{
  \begin{array}{lcl}
\displaystyle{-\frac{3M_{\rm i}}{4\pi^2}\int_{\tau_{\rm UV}}^{\infty}{\rm d}\tau \frac{e^{-\tau M_{\rm i}^2}}{\tau^2}},\qquad\qquad\qquad\qquad\qquad(for\,\,T=0,\,\mu=0)\label{mutoqc}\\
\displaystyle{-\frac{3M_{\rm i}}{\pi^2}\int_{\sqrt{\mu^2-M_{\rm i}^2}}^{+\infty}{\rm d}p\textstyle{\frac{\left[1-{\rm Erf}(\sqrt{M_{\rm i}^2+p^2}\sqrt{\tau_{\rm UV}})\right]p^2}{\sqrt{M_{\rm i}^2+p^2}}}},\,\,\,\,(for\,\,T=0,\,\mu\neq0,\,\,and\,\,M_{\rm i}<\mu)\nonumber\\
\displaystyle{\frac{3M_{\rm i}}{4\pi^2}\left[\textstyle{-M_{\rm i}^2{\rm Ei}(-M_{\rm i}^2\tau_{\rm UV})-\frac{e^{-M_{\rm i}^2\tau_{\rm UV}}}{\tau_{\rm UV}}}\right]},\quad\quad\,\,\,\,\,\,(for\,\,T=0,\,\mu\neq0,\,\,and\,\,M_{\rm i}>\mu)
  \end{array}\right.
\end{eqnarray}
\end{widetext}
where the integral limit $\tau_{\rm UV}$ is a $\Lambda_{\rm UV}$ (ultraviolet cutoff)-related parameter and is defined as $\tau_{\rm UV}=\Lambda_{\rm UV}^{-2}$.

\begin{table}
\caption{Parameter set fixed in our work. The coupling constants $G$ and $K$ have the unit of MeV$^{-2}$ and MeV$^{-5}$, respectively, while the unit of other parameters in this table is MeV.}\label{parameters}
\begin{tabular}{p{0.6cm} p{0.6cm} p{0.7cm} p{1.9cm} p{1.9cm}p{0.6cm}p{0.6cm}}
\hline\hline
$m_{\rm u}$&$\,m_{\rm s}$&$\Lambda_{\rm UV}$&$\qquad G$&$\qquad K$&$M_{\rm u}$&$M_{\rm s}$\\
\hline
3.2&99&1380$\,\,$&$\,1.41\times10^{-6}$&$2.36\times10^{-14}$&194&357\\
3.3&102&1350$\,\,$&$\,1.46\times10^{-6}$&$2.55\times10^{-14}$&195&361\\
3.4&104&1330$\,\,$&$\,1.51\times10^{-6}$&$2.75\times10^{-14}$&197&364\\
3.5&108&1310$\,\,$&$\,1.56\times10^{-6}$&$2.96\times10^{-14}$&198&367\\
3.6&110&1290$\,\,$&$\,1.61\times10^{-6}$&$3.18\times10^{-14}$&199&371\\
3.7&113&1270$\,\,$&$\,1.66\times10^{-6}$&$3.41\times10^{-14}$&200&374\\
3.8&116&1250$\,\,$&$\,1.72\times10^{-6}$&$3.64\times10^{-14}$&202&377\\
3.9&119&1235$\,\,$&$\,1.77\times10^{-6}$&$3.89\times10^{-14}$&203&380\\
4.0&121&1220$\,\,$&$\,1.82\times10^{-6}$&$4.15\times10^{-14}$&204&384\\
4.1&125&1200$\,\,$&$\,1.88\times10^{-6}$&$4.42\times10^{-14}$&205&388\\
\hline\hline
\end{tabular}
\end{table}

Now from the seven parameters present in the above equations five are fitted to reproduce experimental data ($f_{\pi}=92$ MeV, $M_{\pi}=135$ MeV, $M_{K^0}=495$ MeV, $M_{\eta}=548$ MeV, and $M_{\eta '}=958$ MeV) at zero temperature and chemical potential, that is, similar to the process in Ref.~\cite{HATSUDA1994221}, ($M_{\rm u}, \Lambda_{\rm UV}$) to fit ($f_{\pi}, M_{\pi}$), ($M_{\rm s}$, $G$, $K$) to fit ($M_{K^0}, M_{\eta}, M_{\eta '}$), while the parameter $m_{\rm u}$ is fixed before the fitting. Once the above six parameters have been fixed, the value of current quark mass $m_{\rm s}$ can be determined as $m_{\rm s} = M_{\rm s}+4G\langle\bar{s}s\rangle-2K\langle\bar{u}u\rangle^2$. Then for different values of $m_{\rm u}$, the result of parameter sets is listed in Table.~\ref{parameters}. In the latest edition of Review of Particle Physics~\cite{PhysRevD.98.030001}, we notice that the current quark mass $\bar{m}$ and $m_{\rm s}$ are well constrained as $\bar{m}=(m_{\rm u}+m_{\rm d})/2=3.5^{+0.5}_{-0.2}$ MeV, $m_{\rm s}=95^{+9}_{-3}$ MeV. Thus we will choose the parameter sets of $m_{\rm u}=$3.3 MeV and 3.4 MeV to continue the following calculations.

To get the EOS of quark matter at zero temperature and finite chemical potential, we have to deduce the relation of quark density and chemical potential, which is derived as
\begin{eqnarray}
  \rho_{\rm i}(\mu) &=& \langle\psi^+\psi\rangle_{\rm i}\nonumber \\
   &=& -N_{\rm c} \int\frac{{\rm d}^4p}{(2\pi)^4}{\rm tr}\left[iS_{\rm i}\gamma_0\right]\nonumber\\
   &=& 2N_{\rm c}\int\frac{{\rm d}^3p}{(2\pi)^3}\theta(\mu-\sqrt{p^2+M_{\rm i}^2})\nonumber\\
   &=& \left\{
\begin{array}{lcl}
 \frac{1}{\pi^2}(\sqrt{\mu^2-M_{\rm i}^2})^3,             & &\mu>M_{\rm i}\\
  0,                                                & &\mu<M_{\rm i}
   \end{array}
   \right.\label{qnd}
\end{eqnarray}
\begin{figure}
\includegraphics[width=0.47\textwidth]{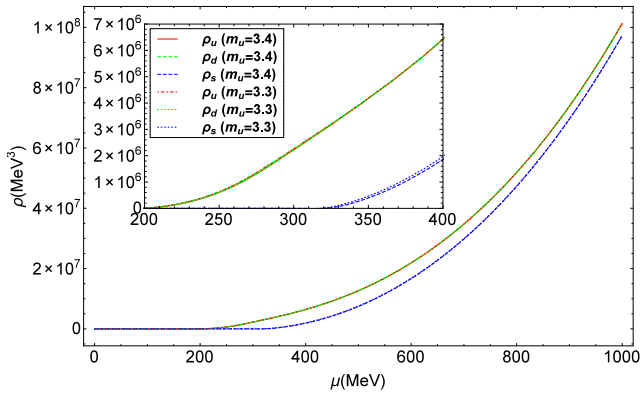}
\caption{Quark number density of $u, d$ and $s$ quark as a function of $\mu$ at $T=0$ with parameters fixed for $m_{\rm u}=$3.3 MeV, 3.4 MeV respectively. Four lines (two red lines and two green lines) nearly coincide and so do the other two blue lines.}
\label{Fig:qnd}
\end{figure}
here "tr" means the trace in Dirac space, and the result is shown in Fig.~\ref{Fig:qnd}. From this figure, we can find that the chemical potential dependence of quark density for $m_{\rm u}=$3.3 MeV and 3.4 MeV are very similar: the critical point $\mu_c=200$ MeV for $u, d$ quark and $\mu_c=320$ MeV for $s$ quark. After these corresponding points, the quark densities start to be nonzero and increase smoothly as the chemical potential increases.

It is noted that the dynamical masses of quarks listed on Table.~\ref{parameters} seem abnormally low compared to what is standard in the literature and the difference between the dynamical mass in vacuum for the strange quark and its critical chemical potential is relatively large ($\sim$40 MeV). In the following, we demonstrate the reasons for that: it is well known that the NJL model is not a renormalizable theory, so we need to use an appropriate regularization to eliminate the ultraviolet (UV) divergence. In the framework of the usual NJL model, three dimensional (3D) momentum cutoff ($\Lambda_{\rm UV}$) regularization is often used to realize that. In this regularization scheme, dynamical quark masses are $M_{\rm u}\sim 350$ MeV, $M_{\rm s}\sim 520$ MeV, which are much larger than the corresponding dynamical quark masses obtained herein, and the chiral phase transition in this case for zero temperature and finite chemical potential is first order. It should be pointed out that for a QCD effective model, $\Lambda_{\rm UV}$ implies the adaptation range of the effective model. Under the normal NJL model framework, the UV cutoff $\Lambda_{\rm UV}$ is about 630 MeV, which means that the NJL model regularized by 3D momentum cutoff cannot be used in principle for physical systems with energy scales greater than $\Lambda_{\rm UV}$= 630 MeV. We know that the energy scale involved in the study of neutron stars is about 1 GeV, thus in this case, we have to abandon the common used 3D momentum cutoff and use PTR instead. This is because PTR is not plagued by the interruption of UV momentum. In this scheme, we can see that the integral limit $\tau_{\rm UV}$ is actually a soft cutoff with the integral variable $\tau$ presenting in the exponential function, and the UV cutoff $\Lambda_{\rm UV}=(\tau_{\rm UV})^{-1/2}$ is set to be larger than 1 GeV by fitting the experimental data. Additionally, the chiral phase transition for T=0 with finite chemical potential is a crossover in PTR. From above we can see that different regularization schemes cause different results. In fact, a certain regularization approach is already employed in the process of parameter fixing. For example, in Refs.~\cite{doi:10.1142/S0217732316500863,Zhao2018,PhysRevD.95.056018}, PTR is also used in NJL model, and the dynamical masses of quarks in these studies ($M_{\rm u}\sim210$ MeV, $M_{\rm s}\sim400$ MeV) are also quite smaller than the usual dynamical quark masses in the normal NJL molel ($M_{\rm u}\sim 350$ MeV, $M_{\rm s}\sim 520$ MeV). In Fig. 1 of the manuscript, we demonstrate the densities of quarks versus the chemical potential for $m_{\rm u}=3.3$ MeV, 3.4 MeV whose corresponding dynamical masses of $s$ quark in vacuum are fixed to be about 360 MeV. The difference between dynamical mass in vacuum and $\mu_C$ for $s$ quark is about 40 MeV. Actually, for other studies in the framework of 2+1 flavors NJL model with PTR such as Refs.~\cite{Zhao2018,PhysRevD.95.056018}, the difference is also large, that is, $\sim$ 40 MeV in Ref.~\cite{Zhao2018} and $\sim$ 80 MeV in Ref.~\cite{PhysRevD.95.056018}.

Considering the internal environment of a hybrid star, we have to take the chemical equilibrium and electric charge neutrality into account,
\begin{equation}\label{constrains}
  \left\{\begin{array}{lcl}
           \mu_{\rm d}=\mu_{\rm u}+\mu_{\rm e}. \\
           \mu_{\rm s}=\mu_{\rm u}+\mu_{\rm e}. \\
           \frac{2}{3}\rho_{\rm u}-\frac{1}{3}\rho_{\rm d}-\frac{1}{3}\rho_{\rm s}-\rho_{\rm e}=0.
         \end{array}\right.
\end{equation}
\begin{figure}
\includegraphics[width=0.47\textwidth]{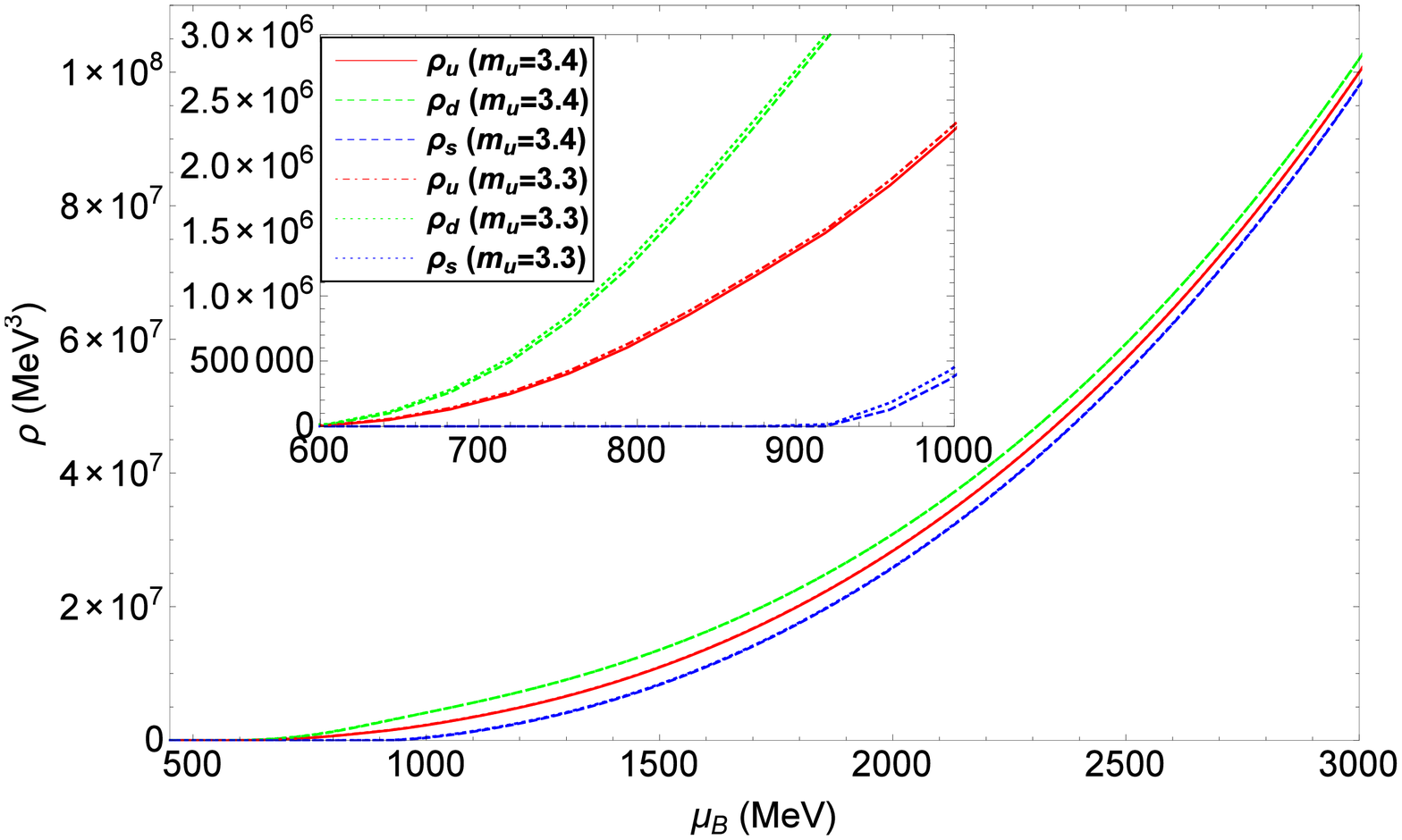}
\caption{Considering the chemical equilibrium and electric charge neutrality of the quark system, density of $u, d$ and $s$ quark with parameters fixed for $m_{\rm u}=$3.3 MeV, 3.4 MeV are shown respectively. Two red lines nearly coincide and so do two green lines and two blue lines.}
\label{Fig:qndb}
\end{figure}

Then we can get the baryon chemical potential dependence of the quark densities, which is presented in Fig.~\ref{Fig:qndb}. As we can see, for a given flavor of quark, the density dependences on baryon chemical potential for $m_{\rm u}=3.3$ MeV and 3.4 MeV are also very similar. The critical baryon chemical potential is $\mu_B^c=600$ MeV for $u$ and $d$ quark, and $\mu_B^c=920$ MeV for $s$ quark. After the corresponding $\mu_B^c$, each density in this picture increases monotonously and smoothly.

According to definition, at zero temperature and finite chemical potential, the EOS of QCD can be written as~\cite{doi:10.1142/S0217751X08040457,PhysRevD.78.054001}
\begin{equation}\label{EOSofQCD}
  P(\mu)=P(\mu=0)+\int_{0}^{\mu}d\mu'\rho(\mu'),
\end{equation}
here $P(\mu=0)$ represents the negative pressure of the vacuum, which is taken as a phenomenological model-dependent parameter. Furthermore, it can reflect the confinement of QCD just like in the MIT bag model. Same to Ref.~\cite{PhysRevD.92.054012}, we regard $P(\mu=0)$ as $-B$ (vacuum bag constant). From Eq.~(\ref{EOSofQCD}), we can deduce that similar behaviors of quark densities between two schemes $m_{\rm u}=3.3$ MeV and 3.4 MeV will result in similar EOSs of quark matter. Thus we will take the scheme of $m_{\rm u}=3.4$ MeV to continue the following study.
After we determine the value of $B$, the energy density can be calculated by~\cite{PhysRevD.86.114028,PhysRevD.51.1989}
\begin{equation}\label{rbedasp}
  \epsilon=-P+\sum_{i}\mu_{\rm i}\rho_{\rm i}.
\end{equation}

\subsection{Hybrid EOS constructed by a three-window modeling}
To get the hybrid EOS with a crossover hadron-quark phase transition, we have to employ a suitable interpolating approach to connect the hadronic EOS and quark EOS. In Refs.~\cite{Masuda01072013,PhysRevC.93.035807,KOJO2016821,PhysRevD.92.054012,PhysRevD.95.056018}, a three-window modeling is adopted. In particular, Refs.~\cite{Masuda01072013,PhysRevC.93.035807,KOJO2016821} employ the $\epsilon-$interpolation in $\epsilon-\rho$ plane or/and $P-$interpolation in $P-\rho$ plane; Refs.~\cite{PhysRevD.92.054012,PhysRevD.95.056018} take $P-$interpolation in $P-\mu$ plane. Just as Ref.~\cite{PhysRevC.93.035807} claims, the three-window modeling is a phenomenological modeling approach. Beyond mere interpolation, different interpolating schemes will have different additional thermodynamic corrections to the interpolated variables, meanwhile, the additional corrections have to preserve the thermodynamic consistency between the variables. In fact, any interpolating approach above is applicable. Although the hybrid EOSs in these three schemes contain different variables, they all satisfy the thermodynamic consistency in the crossover region. As a result, they should match each other in the same plane. In addition, for densities that are very small or very large, the hybrid EOSs will revert to the hadronic EOS or quark EOS, thus matching each other too. In this paper, we will use the same interpolating approach as Refs.~\cite{PhysRevD.92.054012,PhysRevD.95.056018}. By definition, the interpolation function is
\begin{eqnarray}
  P(\mu) &=& P_{\rm H}(\mu)f_-(\mu)+P_{\rm Q}(\mu)f_+(\mu),\,\,\nonumber\\
  f_{\pm}(\mu) &=& \frac{1}{2}(1\pm {\rm tanh}\,(\frac{\mu-\tilde{\mu}}{\Gamma})),\,\,\label{interpolating}
\end{eqnarray}
and the energy density is obtained from the thermodynamic relation
\begin{eqnarray}
  \epsilon(\mu) &=& \epsilon_{\rm H}(\mu)f_-(\mu)+\epsilon_{\rm Q}(\mu)f_+(\mu)+\Delta\epsilon,\,\,\nonumber\\
  \Delta\epsilon &=& \mu(P_{\rm Q}-P_{\rm H})g(\mu),\,\,\label{interpolatingaddition}
\end{eqnarray}
where $P_{\rm H}$ and $P_{\rm Q}$ denote the pressure in hadronic phase and quark phase, respectively. The sigmoid interpolating functions $f_{\pm}$ can realize a smooth DPT in the region of $\tilde{\mu}-\Gamma\lesssim\mu\lesssim\tilde{\mu}+\Gamma$, which is named the window of the function. In this region, hadrons are hybrid with quarks: they coexist and interact strongly. In Eq.~(\ref{interpolatingaddition}), $\Delta\epsilon$ is the additional term that guarantees thermodynamic consistency with $g(\mu)=\frac{2}{\Gamma}(e^{\rm X}+e^{\rm -X})^{-2}$ and ${\rm X}=(\mu-\tilde{\mu})/\Gamma$. From Eq.~(\ref{interpolating}) and Eq.~(\ref{interpolatingaddition}), we can see that there are two parameters in our interpolating procedure: the central baryon chemical potential of the interpolating area $\tilde{\mu}$ and half of the interpolating interval $\Gamma$.

From the study above, we can conclude that the constructed hybrid EOS contains three parameters undetermined totally, i.e. $B$, $\tilde{\mu}$, and $\Gamma$. Thus we can regard our hybrid EOS as a function of these three parameters.
\section{METHODS}\label{two}
In our study, we consider the following five constraints to restrict the EOS of hybrid stars:

(1) The mass constraint from PSR J0348+0432 requires the maximum mass of the neutron star larger than 1.97 $M_{\odot}$~\cite{Antoniadis1233232}.

(2) Because of the uncertainty of $\mu_{\rm deconfinement}$, many studies employ an assumption that $\mu_{\rm deconfinement}\sim\mu_{\rm ChiralRestoration}$~\cite{Masuda01072013,0004-637X-764-1-12,PhysRevD.92.054012}. However, the studies of QCD phase diagram~\cite{PhysRevD.77.114028,0034-4885-74-1-014001} imply that $\mu_{\rm deconfinement}>\mu_{\rm ChiralRestoration}\sim1$ GeV at zero temperature with finite chemical potential. Thus in this paper, we take a relatively loose constraint that $\tilde{\mu}-\Gamma \geq 1$ GeV in the hybrid construction.

(3) The latest update of the source properties for GW170817 from LIGO and Virgo collaborations~\cite{Abbott:2018wiz} demonstrates that the dimensionless combined tidal deformability $\tilde{\Lambda}$ has a considerable change compared with the former observable, that is, $\tilde{\Lambda}\sim280^{+490}_{-190}$ for the case of symmetric $90\%$ credible interval and $\tilde{\Lambda}\sim280^{+410}_{-230}$ for the case of highest posterior density (HPD) $90\%$ credible interval. The definition of it is shown in the following,
\begin{equation}\label{combinedtd}
\noindent\tilde{\Lambda}=\frac{16}{13}\frac{(M_1+12M_2)M_1^4\Lambda_1+(M_2+12M_1)M_2^4\Lambda_2}{(M_1+M_2)^5}.
\end{equation}
Here $\Lambda_1, \Lambda_2$ are the deformability of the two members of BNS, and $M_1, M_2$ are the corresponding gravitational masses, respectively. The detailed calculation method of $\Lambda$ and its dependence on $M$ can be found in Ref.~\cite{PhysRevD.81.123016}. With the additional waveform model SEOBNRT, the chirp mass $\mathcal{M}=(M_1 M_2)^{3/5}(M_1+M_2)^{-1/5}$ is fixed to $1.186\pm0.0001 M_{\odot}$ (This value determines the relation of $M_1$ and $M_2$).

(4) The stability in interpolating between the quark EOS and hadronic EOS demands ${\rm d}P/{\rm d}\rho>0$, and it is very restrictive to the interpolated EOS~\cite{PhysRevC.93.035807}. Actually, ${\rm d}P/{\rm d}\rho$ is relevant to the sound velocity of the system which is defined as $v=\sqrt{{\rm d}P/{\rm d}\epsilon}$. Via Eqs.~(\ref{EOSofQCD}) and~(\ref{rbedasp}), we can derive that
$v^2={\rm d}P/{\rm d}\epsilon={\rm d}P/(-{\rm d}P+\rho{\rm d}\mu+\mu{\rm d}\rho)=1/\mu\cdot{\rm d}P/{\rm d}\rho$. Thus this constraint is equivalent to $v^2>0$.

(5) The stability of the hybrid star with a maximum mass requires $\mu_{\rm C}>\mu_{\rm BE}$, where $\mu_{\rm C}$ is the baryon chemical potential in the center of the star, and $\mu_{\rm BE}$ represents the baryon chemical potential of the intersection between quark binding energy and hadronic binding energy. For $\mu<\mu_{\rm BE}$, the hadronic matter is more stable with a lower binding energy than quark matter; but for $\mu>\mu_{\rm BE}$, the inverse is true. Therefore, $\mu_{\rm C}>\mu_{\rm BE}$ should be satisfied to forbid the quark matter decaying into the hadronic matter in the center of the heaviest star. Only in this way, the deconfined regime (pure or mixed phase) can be achieved, and the hybrid star not the pure neutron star (a scenario which we find ruled out by the latest observation data from GW170817 pertaining tidal deformability) can exist.
\section{RESULTS}\label{three}
\begin{figure}
\includegraphics[width=0.47\textwidth]{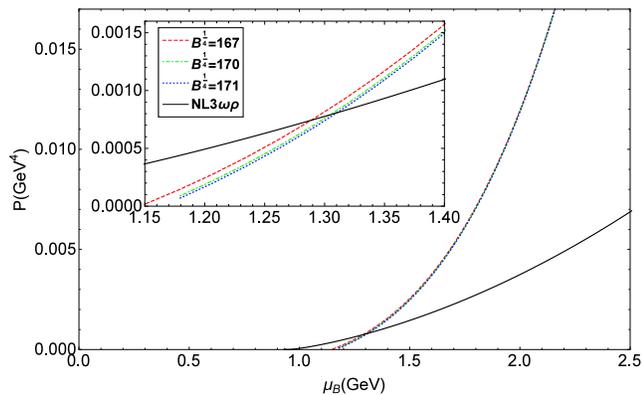}
\caption{Comparison of quark EOSs and hadronic EOS. The black solid line is the NL3$\omega\rho$ EOS while the red dashed line, the green dot-dashed line and the blue dotted line are the quark EOSs with $B^{\frac{1}{4}}=$167 MeV, 170 MeV, and 171 MeV respectively.}
\label{Fig:pmurelation}
\end{figure}

\begin{figure}
\includegraphics[width=0.47\textwidth]{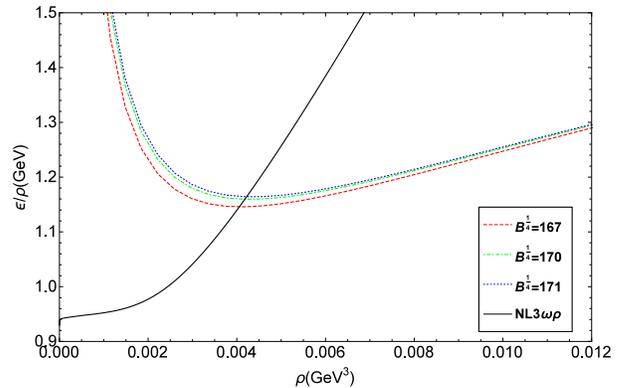}
\caption{Comparison of binding energy of hadrons and quarks. The black solid line is for the NL3$\omega\rho$ EOS while the red dashed line, the green dot-dashed line and the blue dotted line are for the quark EOSs with $B^{\frac{1}{4}}=$167 MeV, 170 MeV, and 171 MeV respectively.}
\label{Fig:bindingenergy}
\end{figure}

We choose $B^{\frac{1}{4}}=167$, 170, and 171 MeV as three representative values to compare the EOSs of quark matter and hadronic matter, and the result is shown in Fig.~\ref{Fig:pmurelation}. We can see that for a larger value of $B^{\frac{1}{4}}$, the pressure is also larger for the same $\mu_B$, but quark EOSs do not differ too much in these three cases. The intersections of quark EOSs and the NL3$\omega\rho$ EOS are located at around $\mu_B=1.3$ GeV. Then we calculate the binding energy $\epsilon/\rho$ of quarks for the three representative values of $B^{\frac{1}{4}}$, and compare the result with that of NL3$\omega\rho$ model, which is shown in Fig.~\ref{Fig:bindingenergy}. From this figure, we can find that for a certain density, as $B^{\frac{1}{4}}$ increases, the binding energy also increases, and the intersections of quark binding energy and hadronic binding energy are close to $\rho=0.004$ GeV$^3$. In the left side domain of the intersection, the binding energy of hadrons is smaller than that of quarks, indicating hadrons are more stable than quarks. However, in the right side domain of the intersection, conversely, quarks are more stable with a smaller binding energy than hadrons.

\begin{figure}
\centering
\subfigure[]{
\centering
\includegraphics[width=0.4\textwidth]{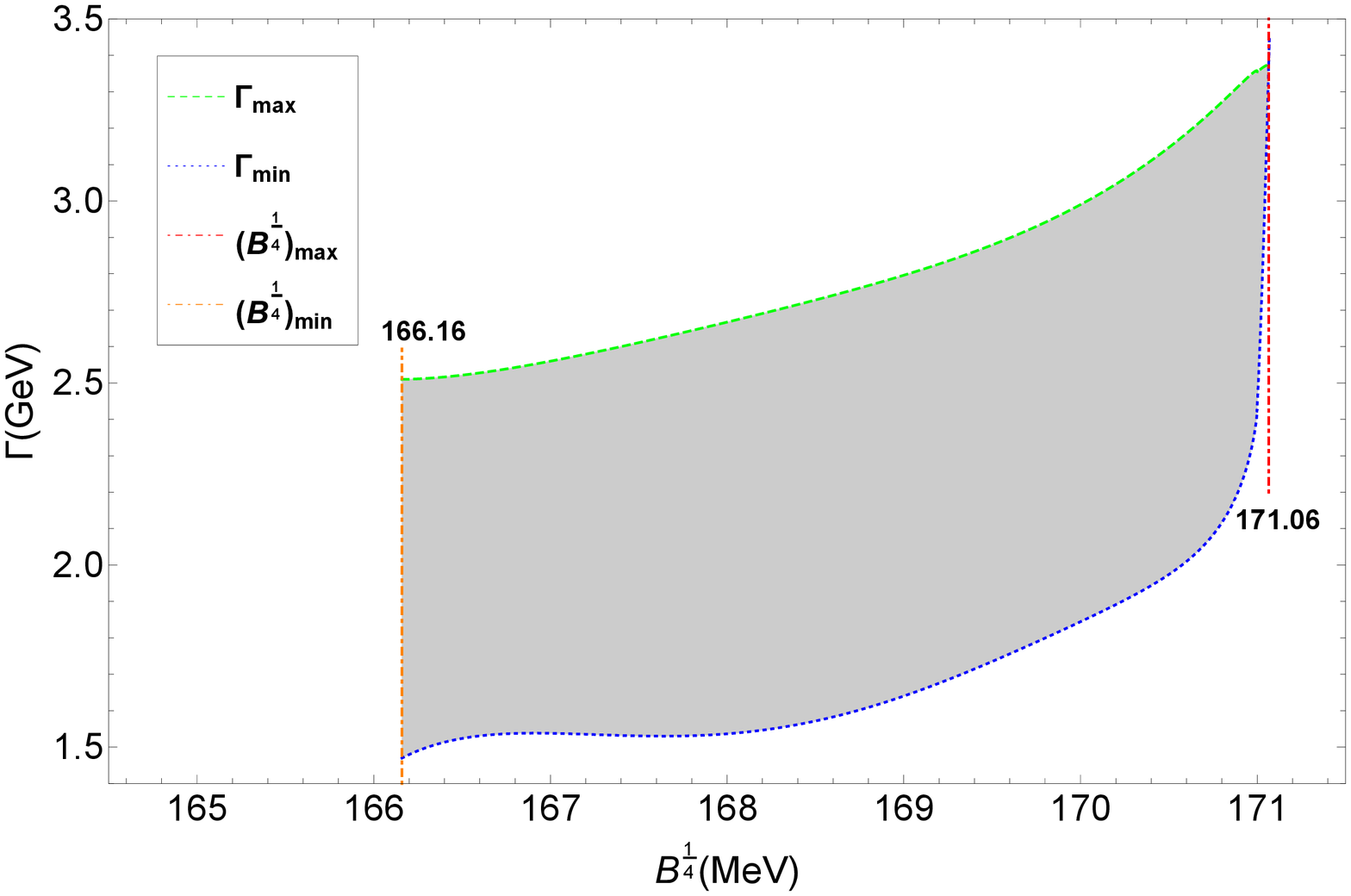}
}
\subfigure[]{
\centering
\includegraphics[width=0.4\textwidth]{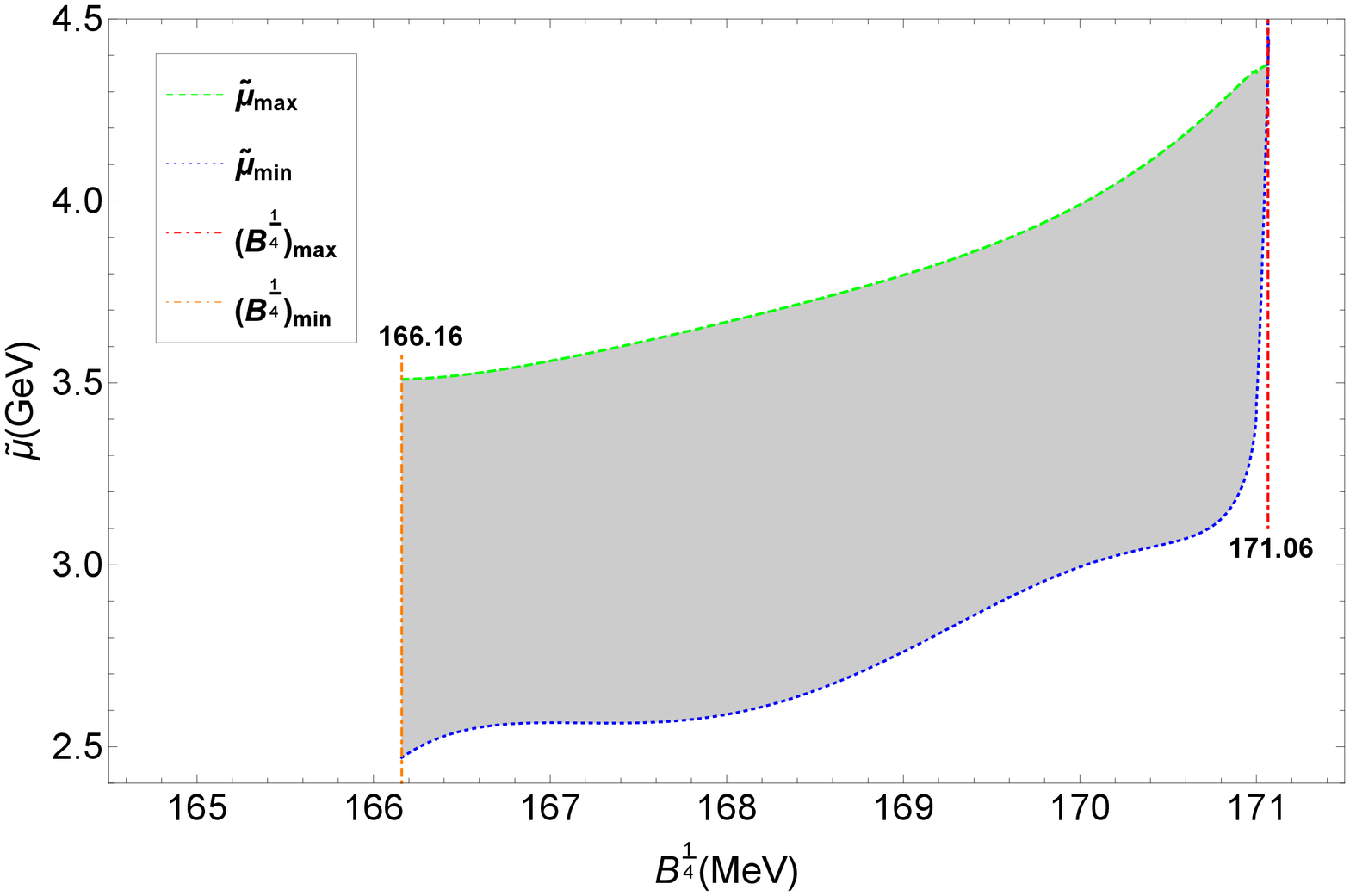}
}
\caption{Constraints on parameter set ($B^{\frac{1}{4}}, \tilde{\mu}, \Gamma$). The gray shaded region is the allowed space for the sub parameter set (a) ($B^{\frac{1}{4}}, \Gamma$), and (b) ($B^{\frac{1}{4}}, \tilde{\mu}$) respectively with five constraints considered in Sec.~\ref{two}}
\label{Fig:bmugamma}
\end{figure}

Then we extend our study to various hybrid EOS models with different parameter sets of ($B^{\frac{1}{4}}, \tilde{\mu}, \Gamma$). With the five constraints considered in Sec.~\ref{two}, it is possible for us to get reasonable choice of the parameter set. Firstly, we consider the constraint on ($B^{\frac{1}{4}}, \Gamma$) and ($B^{\frac{1}{4}}, \tilde{\mu}$) with an appropriate value of $\tilde{\mu}$ and $\Gamma$, respectively. In other words, supposing the allowed space of ($B^{\frac{1}{4}}, \tilde{\mu}, \Gamma$) forming a three-dimensional image, we extract its projection on $\Gamma$-$B^{\frac{1}{4}}$ plain and $\tilde{\mu}$-$B^{\frac{1}{4}}$ plain, respectively. The result is presented in Fig.~\ref{Fig:bmugamma}. From the graph, we can see that for the hybrid EOS, the range of $B^{\frac{1}{4}}$ is restricted to (166.16, 171.06) MeV, and as $B^{\frac{1}{4}}$ increases, the allowed intervals of $\Gamma$ and $\tilde{\mu}$ reduce with both the upper limit and lower limit rising. In particular, for $B^{\frac{1}{4}}=166.16$ MeV, the range of $\Gamma$ and $\tilde{\mu}$ is (1.47, 2.51) GeV and (2.47, 3.51) GeV respectively; but for $B^{\frac{1}{4}}=171.06$ MeV, $\Gamma$ and $\tilde{\mu}$ is constrained to 3.37 and 4.37 GeV respectively.

\begin{figure}[h!]
\centering
\subfigure[]{
\label{Fig:b167mugamma}
\includegraphics[width=0.4\textwidth]{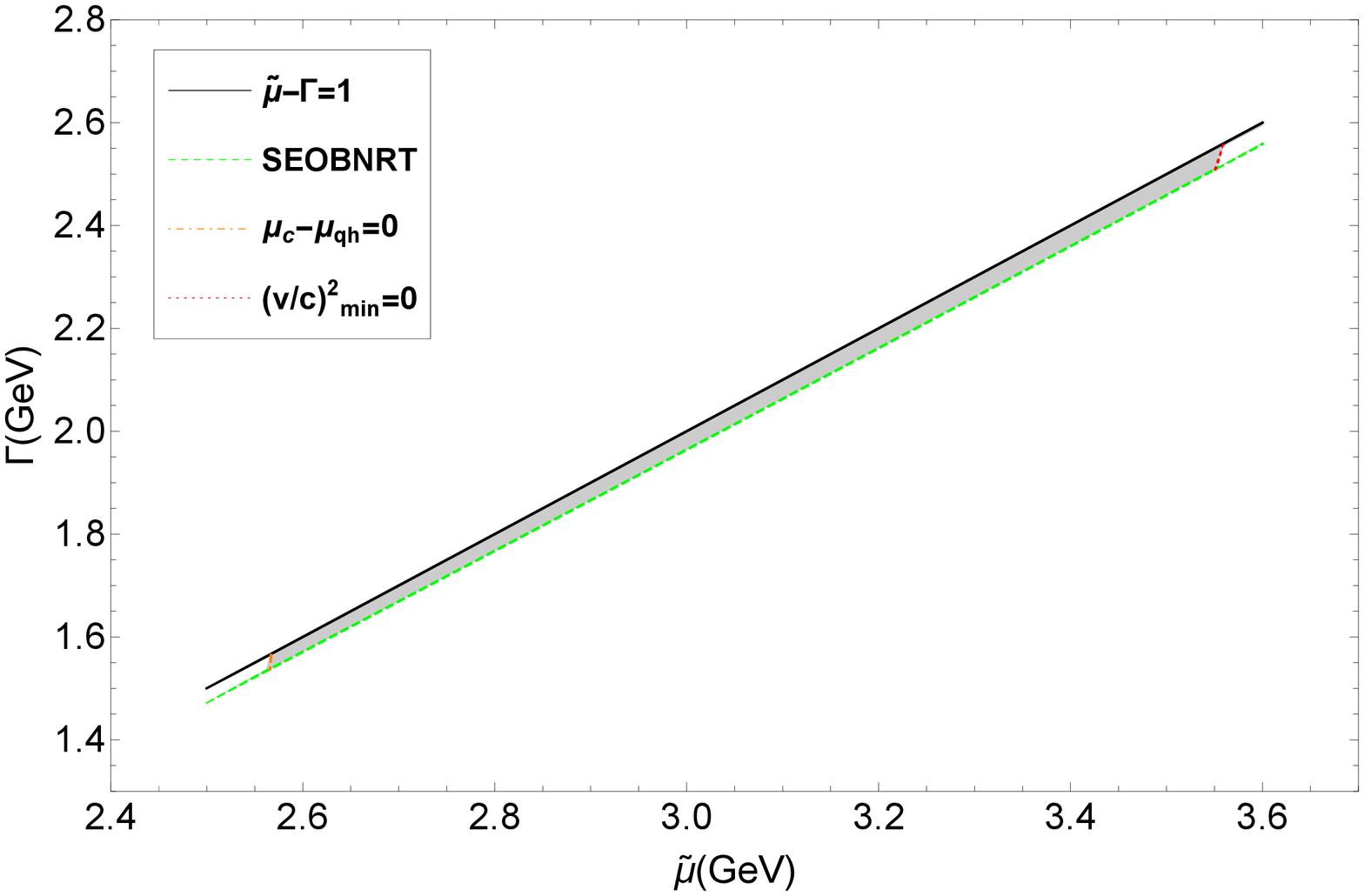}}
\subfigure[]{
\label{Fig:b170mugamma}
\includegraphics[width=0.4\textwidth]{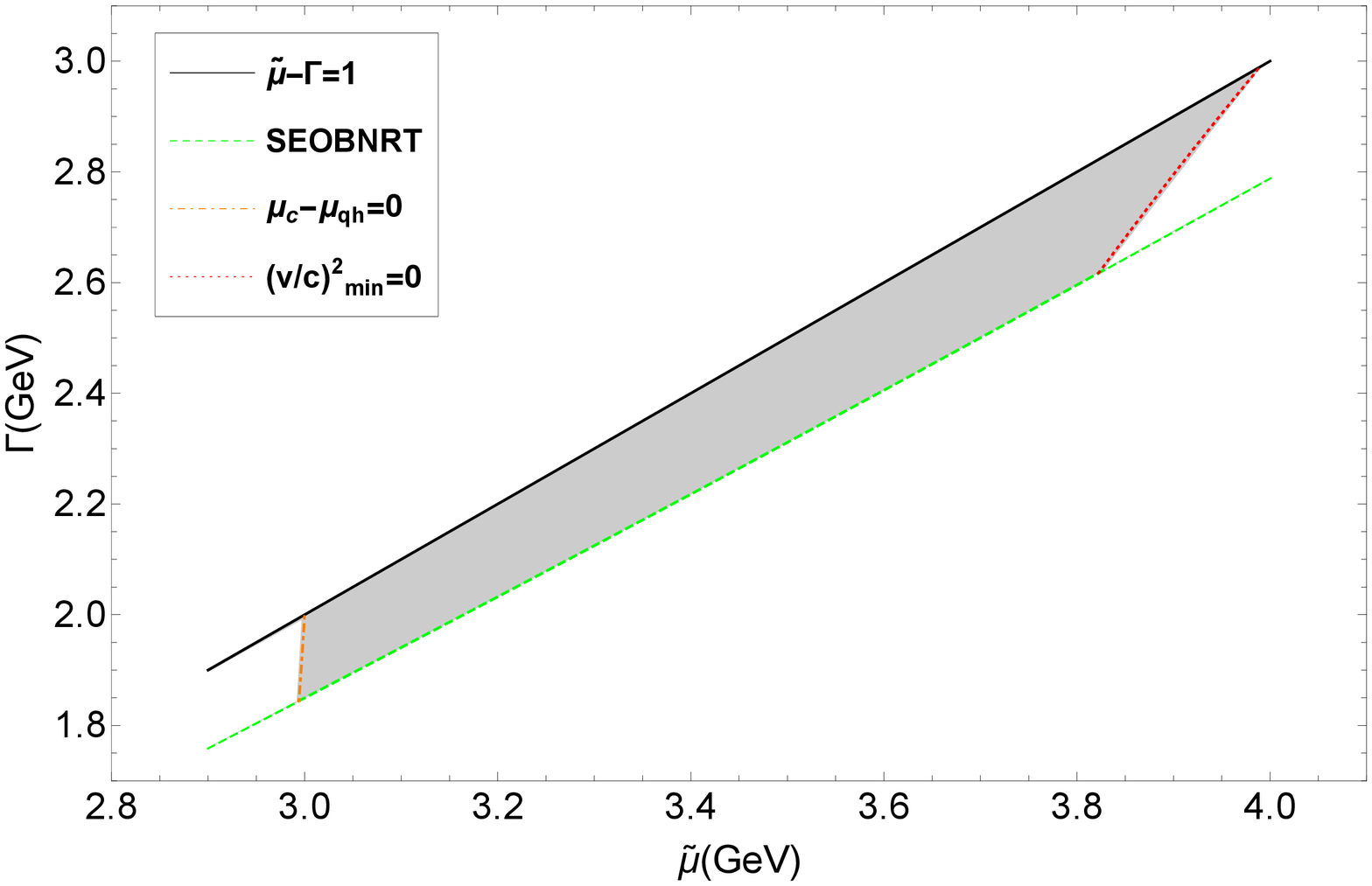}}
\subfigure[]{
\label{Fig:b171mugamma}
\includegraphics[width=0.4\textwidth]{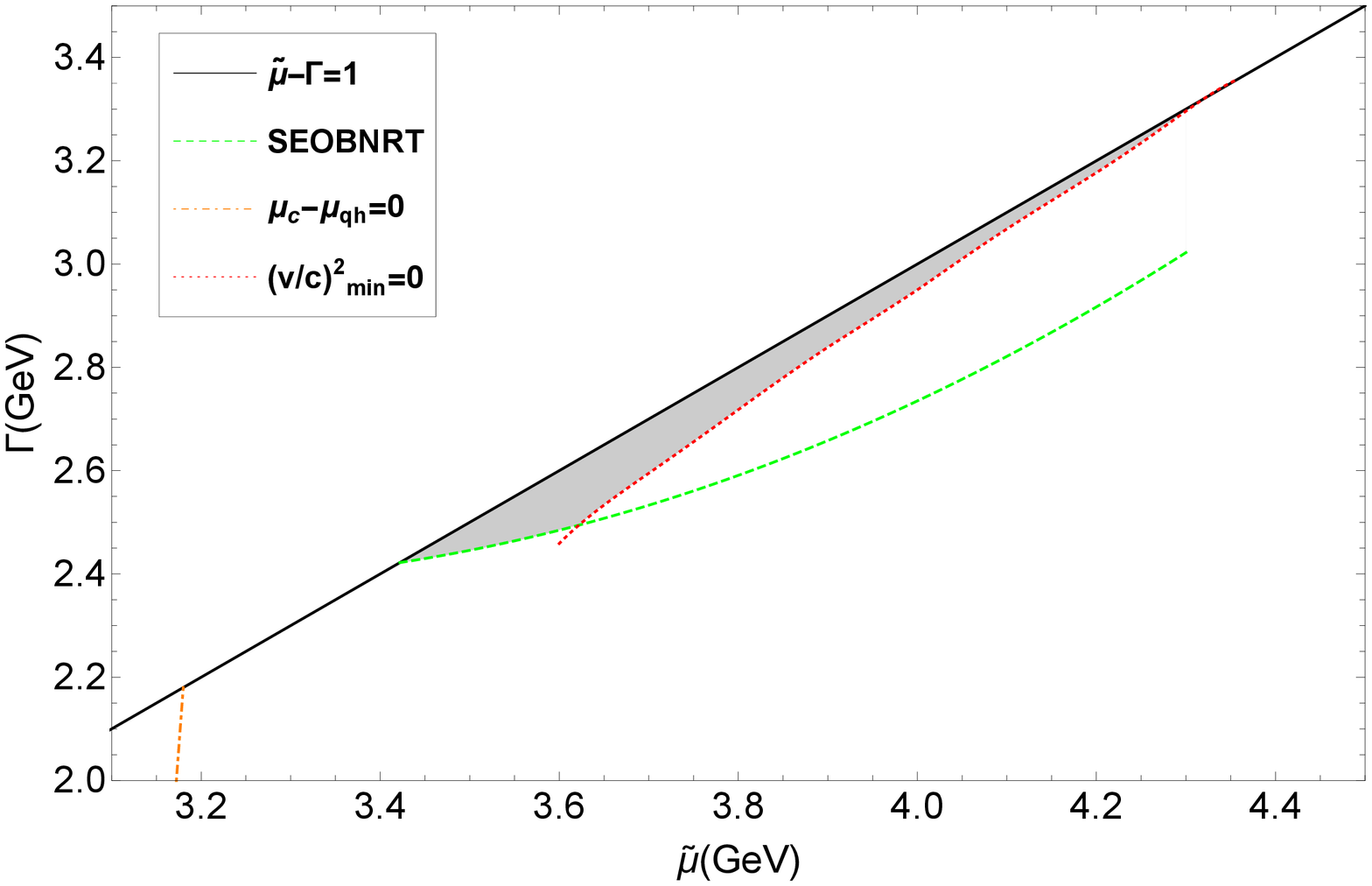}}
\caption{Constraints on sub parameter set ($\tilde{\mu}, \Gamma$) with (a) $B^{\frac{1}{4}}=167$ MeV, (b) $B^{\frac{1}{4}}=170$ MeV, and (c) $B^{\frac{1}{4}}=171$ MeV respectively. The gray shaded region is the allowed parameter space for these three cases. The black solid line, green dashed line, red dotted line, and orange dot-dashed line correspond to the constraints (2), (3), (4), and (5) in Sec.~\ref{two}, respectively. The mass constraint (1) does not appear in these graphs because this constraint is relatively loose. When $\tilde{\mu}>1.95$ GeV, the maximum masses of hybrid stars constructed by the hybrid EOS are already well beyond 1.97 $M_{\odot}$.}
\label{Fig:mugamma}
\end{figure}

Generally, if we want to get the constraint on sub parameter set ($\tilde{\mu}, \Gamma$), $B^{\frac{1}{4}}$ should be fixed to a certain value. In the following, we will study it for three representative schemes, i.e. $B^{\frac{1}{4}}=167$ MeV, 170 MeV, and 171 MeV, and the result is shown in Fig.~\ref{Fig:mugamma}. From the comparison of the three subgraphs (a), (b), and (c), we can see that the area of allowed parameter space of ($\tilde{\mu}, \Gamma$) experiences expansion and then narrowing as $B^{\frac{1}{4}}$ increases. For $B^{\frac{1}{4}}=167$ MeV, the allowed region is long and narrow with $\tilde{\mu}\in(2.57, 3.56)$ GeV and $\Gamma\in(1.54, 2.56)$ GeV. In addition, the longitudinal distance of $\tilde{\mu}-\Gamma=$1 line and SEOBNRT line is about 0.03 GeV for $\tilde{\mu}=2.5$ GeV; while for $\tilde{\mu}=3.6$ GeV, the distance is about 0.04 GeV. For $B^{\frac{1}{4}}=170$ MeV, the allowed space is larger than that of $B^{\frac{1}{4}}=167$ MeV with $\tilde{\mu}$ constrained to (2.99, 3.99) GeV and $\Gamma$ constrained to (1.85, 2.99) GeV. The longitudinal distance of $\tilde{\mu}-\Gamma=$1 line and SEOBNRT line is about 0.14 GeV here for $\tilde{\mu}=2.99$ GeV, and 0.22 GeV for $\tilde{\mu}=4$ GeV. For $B^{\frac{1}{4}}=171$ MeV, the area of the allowed region reduces compared to $B^{\frac{1}{4}}=170$ MeV, that is, (3.42, 4.36) GeV and (2.42, 3.36) GeV for the range of $\tilde{\mu}$ and $\Gamma$, respectively. And the longitudinal distance of the constraind area expands and then narrows as $\tilde{\mu}$ increases. In fact, from our calculations, we also find the following two trends with $B^{\frac{1}{4}}$ increasing: 1, the intersection of $\tilde{\mu}-\Gamma=1$ line and SEOBNRT line as well as the intersection of $\tilde{\mu}-\Gamma=1$ line and $(v/c)^2_{min}=0$ line move to the right side of $\tilde{\mu}-\Gamma$ plane. 2, the SEOBNRT line is trending to the direction of $\tilde{\mu}$ axis but $(v/c)^2_{min}=0$ line is trending to $\tilde{\mu}-\Gamma=1$ line. It should be mentioned that the mass constraint is not shown in Fig.~\ref{Fig:mugamma}, because the mass constraint here is relatively loose compared with the other four constraints.

\begin{figure}
\includegraphics[width=0.47\textwidth]{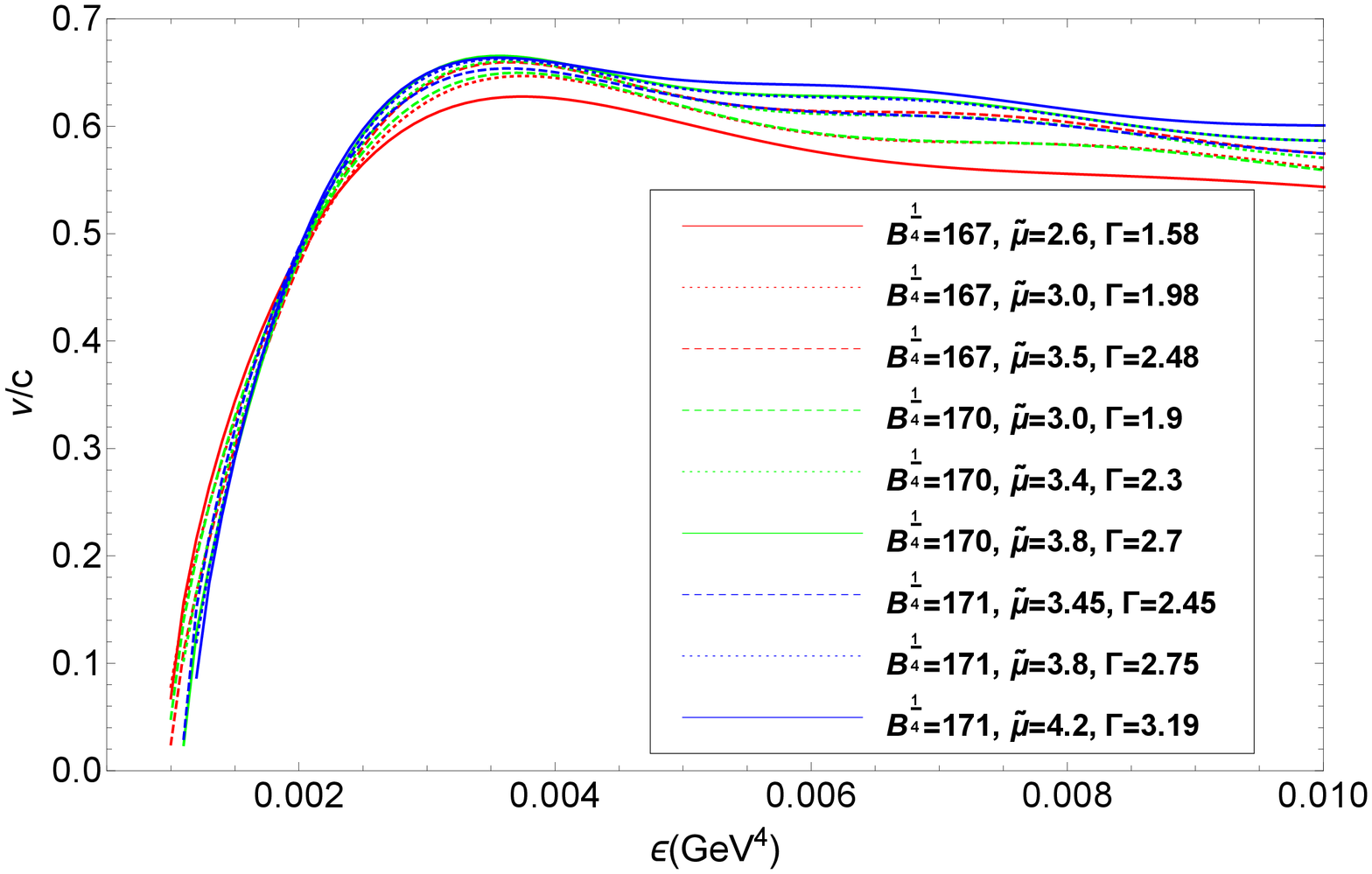}
\caption{The sound velocities of the nine representative hybrid EOSs with parameter set of ($B^{\frac{1}{4}}, \tilde{\mu}, \Gamma)=$(167, 2.6, 1.58), (167, 3.0, 1.98), (167, 3.5, 2.48), (170, 3.0, 1.9), (170, 3.4, 2.3), (170, 3.8, 2.7), (171, 3.45, 2.45), (171, 3.8, 2.75), and (171, 4.2, 3.19), corresponding to the red solid line, red dotted line, red dashed line, green dashed line, green dotted line, green solid line, blue dashed line, blue dotted line, and blue solid line respectively.}
\label{Fig:velocities}
\end{figure}

\begin{figure}
\includegraphics[width=0.47\textwidth]{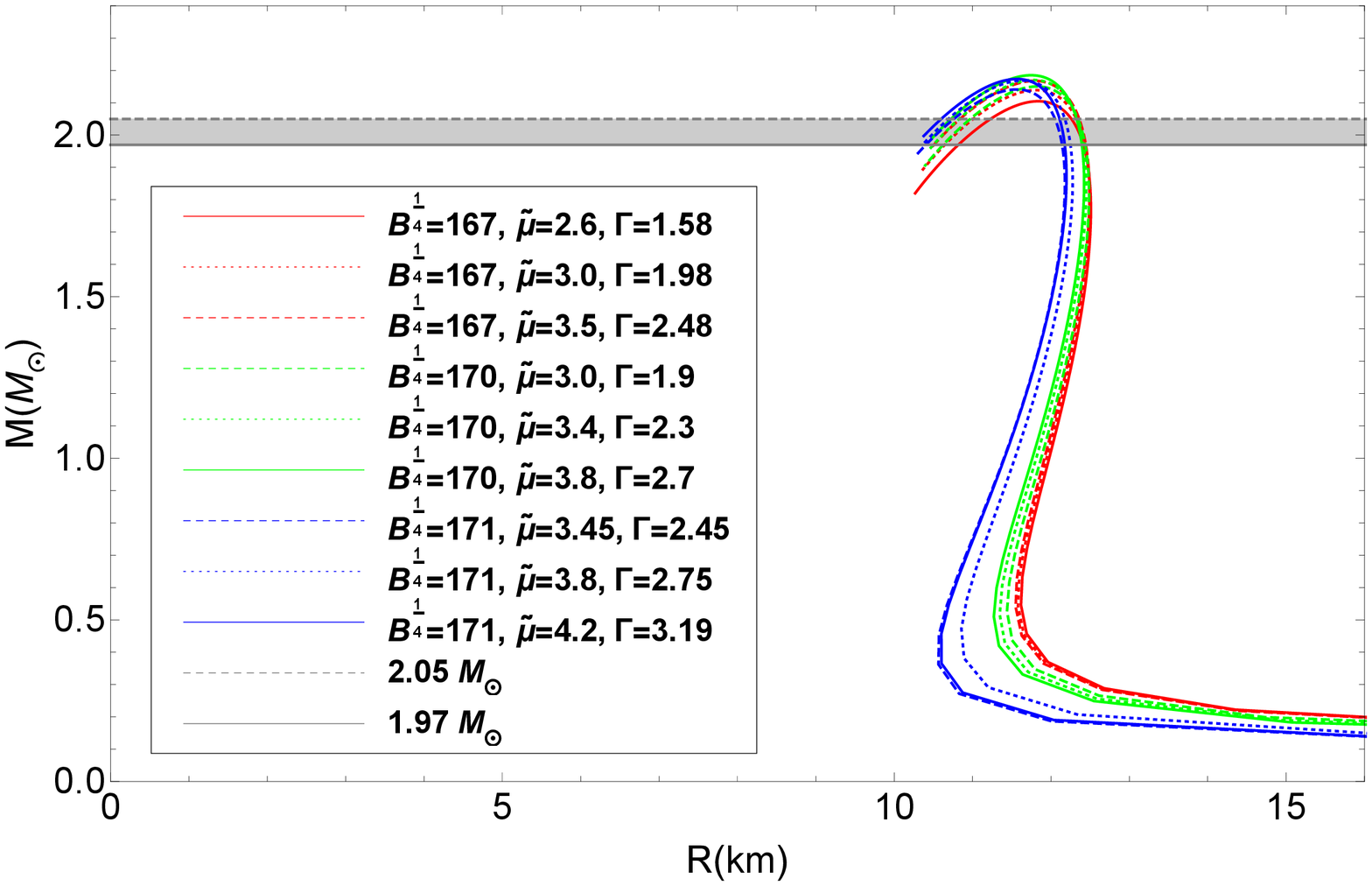}
\caption{The $M-R$ relation of the hybrid stars constructed by the nine representative hybrid EOSs with parameter set of ($B^{\frac{1}{4}}, \tilde{\mu}, \Gamma)=$(167, 2.6, 1.58), (167, 3.0, 1.98), (167, 3.5, 2.48), (170, 3.0, 1.9), (170, 3.4, 2.3), (170, 3.8, 2.7), (171, 3.45, 2.45), (171, 3.8, 2.75), and (171, 4.2, 3.19), corresponding to the red solid line, red dotted line, red dashed line, green dashed line, green dotted line, green solid line, blue dashed line, blue dotted line, and blue solid line respectively. The gray shaded area represents the mass constraint of PSR J0348+0432.}
\label{Fig:mrrelation}
\end{figure}

\begin{figure}
\includegraphics[width=0.47\textwidth]{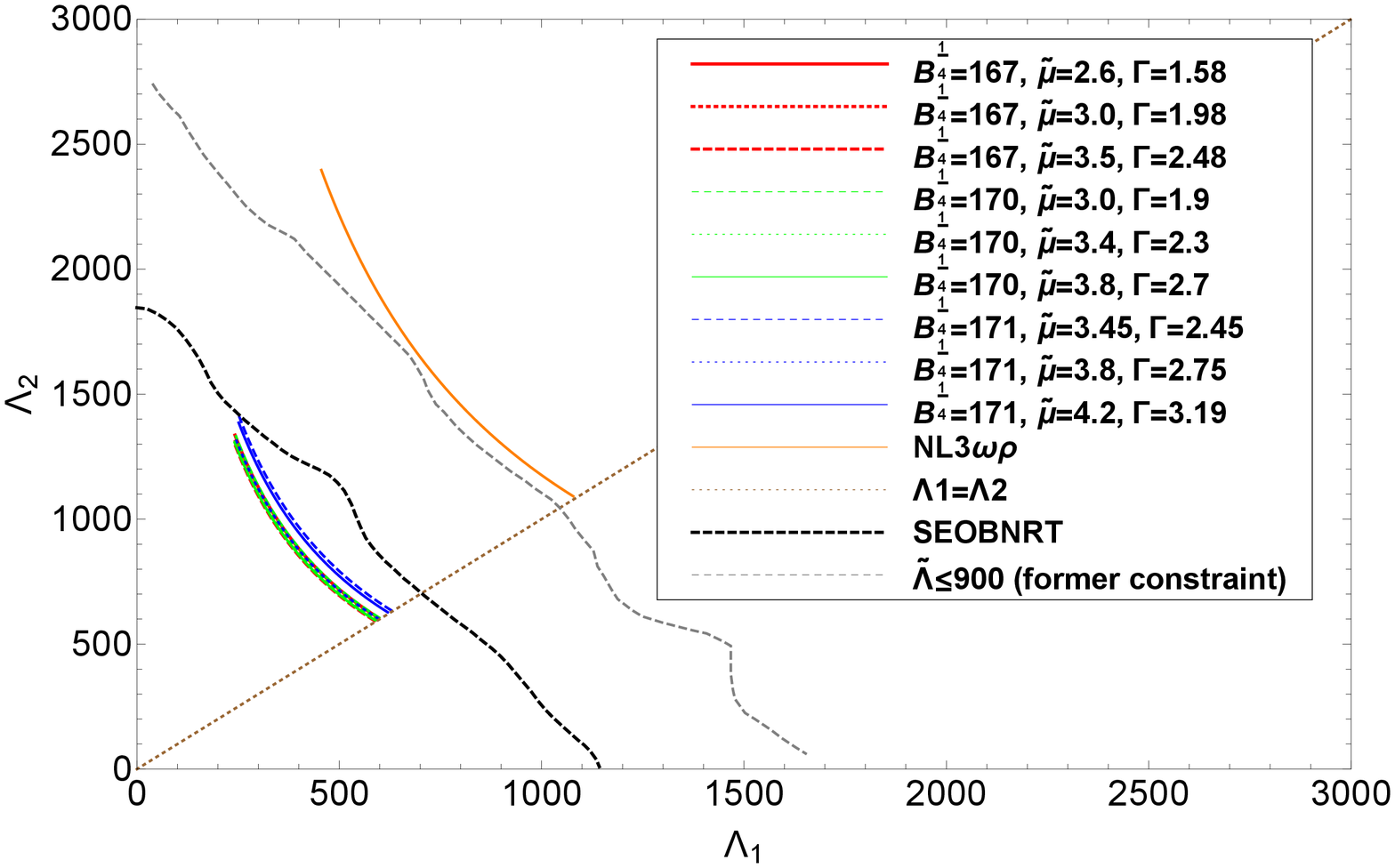}
\caption{Comparison of the tidal deformability of hybrid stars constructed by the nine representative hybrid EOSs with parameter set of ($B^{\frac{1}{4}}, \tilde{\mu}, \Gamma)=$(167, 2.6, 1.58), (167, 3.0, 1.98), (167, 3.5, 2.48), (170, 3.0, 1.9), (170, 3.4, 2.3), (170, 3.8, 2.7), (171, 3.45, 2.45), (171, 3.8, 2.75), and (171, 4.2, 3.19), corresponding to the red solid line, red dotted line, red dashed line, green dashed line, green dotted line, green solid line, blue dashed line, blue dotted line, and blue solid line respectively. For $B^{\frac{1}{4}}=167$ and 170 MeV, the tidal deformability is very similar. The orange solid line represents the tidal deformability ($\Lambda_1, \Lambda_2$) calculated by NL3$\omega\rho$ EOS. Both the black dashed line and gray dashed line are the $90\%$ posterior probability enclosed inside for the low spin prior case in GW170817. The difference is that the gray one represents the former prediction and the black one is the recent prediction in the light of the additional waveform model SEOBNRT. The brown dotted line indicates the $\Lambda_1=\Lambda_2$ boundary.}
\label{Fig:lambdacompare}
\end{figure}

For a more detailed demonstration of the properties of hybrid EOSs with the parameter set in the constrained region of Fig.~\ref{Fig:mugamma}, we will choose three representative points of ($\tilde{\mu}, \Gamma$) for each of the scheme: $B^{\frac{1}{4}}=167$ MeV, 170 MeV, and 171 MeV, to get nine hybrid EOSs. And then we calculate the corresponding sound velocities, $M-R$ relation and tidal deformability ($\Lambda_1, \Lambda_2$), which are shown in Fig.~\ref{Fig:velocities}, Fig.~\ref{Fig:mrrelation} and Fig.~\ref{Fig:lambdacompare} respectively. From Fig.~\ref{Fig:velocities}, we can see that all sound velocities of the hybrid stars are smaller than 0.7 times speed of light, demonstrating the rationality of the hybrid EOSs. In Fig.~\ref{Fig:mrrelation}, the maximum gravitational masses of HSs are from 2.10 $M_{\odot}$ to 2.19 $M_{\odot}$ with a radius from 11.99 km to 12.13 km, well beyond the mass constraint of 1.97 $M_{\odot}$. And the radius of the hybrid stars with a mass of 1.4 $M_{\odot}$ is from 11.90 to 12.18 km. The detailed information of HSs based on these nine hybrid EOSs is listed in Table.~\ref{ninerepresentative}. According to the chirp mass prediction from SEOBNRT, the mass of two stars in the BNS is calculated to be 1.17 $M_{\odot}$-1.36 $M_{\odot}$ and 1.36 $M_{\odot}$-1.59 $M_{\odot}$, respectively. Therefore, we also present the corresponding central baryon chemical potential $\mu_C$ for 1.17 $M_{\odot}$, 1.36 $M_{\odot}$, and 1.59 $M_{\odot}$ in this table. We can see that the value of $\mu_C(1.17)$ in each hybrid EOS is larger than the corresponding $\tilde{\mu}-\Gamma$, i.e., the starting point of DPT in our hybrid EOS, thus suggesting that both two stars of BNS from GW170817 can be HSs shown in Table.~\ref{ninerepresentative}. In addition, nine values of $\mu_C$ in this table are all located in their corresponding interpolating window, namely, the phase transition region, demonstrating that the heaviest star constructed by our hybrid EOS does not have a pure quark core but a mixed-phase inside. The combined dimensionless tidal deformability $\tilde{\Lambda}$ with a flat prior (symmetric/HPD) are also shown in Table.~\ref{ninerepresentative} whose values are all in the region of $90\%$ credible interval predicted by SEOBNRT. In Fig.~\ref{Fig:lambdacompare}, we can see that the constraint for tidal deformability pairs $\Lambda_1$ and $\Lambda_2$ from SEOBNRT shrinks significantly compared to the former. Although the relation of $\Lambda_1$ and $\Lambda_2$ for NL3$\omega\rho$ EOS is very close to the former constraint, it is far beyond the recent prediction of SEOBNRT. Different from that, the results from the nine representative hybrid EOSs are all in accordance with the constraint. Among them, the hybrid EOSs with the schemes of $B^{\frac{1}{4}}=$167 and 170 give very similar tidal deformability parameter.
\begin{widetext}
\begin{center}
\begin{table}
\caption{Some quantities of HSs corresponding to the nine representative hybrid EOSs: maximum gravitational mass $M_{\rm max}$, radius $R_m$, central baryon chemical potential $\mu_C$, radius of 1.4$M_{\odot}$ star $R(1.4)$, central baryon chemical potential of 1.17$M_{\odot}$ star $\mu_C(1.17)$, central baryon chemical potential of 1.36$M_{\odot}$ star $\mu_C(1.36)$, central baryon chemical potential of 1.59$M_{\odot}$ star $\mu_C(1.59)$, and the combined dimensionless tidal deformability $\tilde{\Lambda}$ with flat prior (symmetric/HPD).}\label{ninerepresentative}
\begin{tabular}{p{1.0cm} p{1.0cm} p{1.0cm} p{1.0cm} p{1.0cm}p{1.0cm}p{1.2cm}p{1.5cm}p{1.5cm}p{1.5cm}p{2.7cm}}
\hline\hline
$B^{\frac{1}{4}}$&$\tilde{\mu}$&$\Gamma$&$M_{\rm max}$&$R_m$&$\mu_C$&$R(1.4)$&$\mu_C(1.17)$&$\mu_C(1.36)$&$\mu_C(1.59)$&$\quad\quad\quad\tilde{\Lambda}$\\
$[{\rm MeV}]$&$[{\rm GeV}]$&$[{\rm GeV}]$&$[M_{\odot}]$&[km]&$[{\rm GeV}]$&[km]&$[{\rm GeV}]$&$[{\rm GeV}]$&$[{\rm GeV}]$&(symmetric/HPD)\\
\hline
\multirow{3}{*}{167}&2.6&1.58&2.10&12.13&1.60&12.15&1.15&1.18&1.23&$\quad\quad$601/570\\
                    &3.0&1.98&2.14&12.00&1.63&12.00&1.15&1.18&1.23&$\quad\quad$595/565\\
                    &3.5&2.48&2.17&12.07&1.65&12.09&1.15&1.18&1.23&$\quad\quad$590/561\\
                    \hline
\multirow{3}{*}{170}&3.0&1.90&2.15&11.99&1.63&11.90&1.15&1.19&1.23&$\quad\quad$602/571\\
                    &3.4&2.30&2.17&12.00&1.65&11.99&1.15&1.19&1.24&$\quad\quad$595/565\\
                    &3.8&2.70&2.19&12.01&1.67&12.00&1.16&1.19&1.24&$\quad\quad$592/561\\
                    \hline
\multirow{3}{*}{171}&3.45&2.45&2.14&12.00&1.66&12.00&1.16&1.20&1.25&$\quad\quad$632/600\\
                    &3.8&2.75&2.17&12.03&1.67&12.18&1.16&1.19&1.24&$\quad\quad$599/568\\
                    &4.2&3.19&2.17&12.00&1.68&12.00&1.16&1.19&1.25&$\quad\quad$624/592\\
\hline\hline
\end{tabular}
\end{table}
\end{center}
\end{widetext}

\section{SUMMARY AND DISCUSSION}\label{four}
In this paper, we try to use the constraint of the additional waveform model SEOBNRT on tidal deformability from the latest GW170817 source properties~\cite{Abbott:2018wiz} to restrict the hybrid EOS constructed by a smooth three-window interpolating approach on $P-\mu$ plain~\cite{PhysRevD.92.054012,PhysRevD.95.056018} between hadronic phase and quark phase. The quark matter is described by 2+1 flavors NJL model and the hadronic matter is characterized by RMF NL3$\omega\rho$ model~\cite{PhysRevLett.86.5647,PhysRevC.94.035804}. In 2+1 flavors NJL model, there are seven model parameters and five of them can be fixed by fitting five experimental data if the other two ($m_{\rm u}$ and $m_{\rm s}$) are determined. To satisfy the prediction of these two parameters from the recent study~\cite{PhysRevD.98.030001}, we choose two sets of parameters within $m_{\rm u}=$3.3 MeV and 3.4 MeV respectively to continue the following calculation but to find the quark densities under these two schemes are very similar, which can spontaneously cause a similarity between their corresponding EOSs. Thus the parameter set within $m_{\rm u}=$3.4 MeV is set as the representative one to participate in our calculations. It is noteworthy that three parameters are still free in the hybrid EOS, i.e., $B^{\frac{1}{4}}$ from the quark EOS, $\tilde{\mu}$ and $\Gamma$ from the interpolating process.

Then by the constraint of SEOBNRT, the mass prediction from PSR J0348+0432~\cite{Antoniadis1233232}, the studies of hadron-quark transition in Refs.~\cite{PhysRevD.77.114028,0034-4885-74-1-014001} implying that $\mu_{\rm deconfinement}>\mu_{\rm ChiralRestoration}\sim1$ GeV at zero temperature with finite chemical potential, the stability of hybrid EOS~\cite{PhysRevC.93.035807}, and the stability of the heaviest HS, we restrict the sub parameter set ($B^{\frac{1}{4}}$, $\tilde{\mu}$) and ($B^{\frac{1}{4}}$, $\Gamma$) to a reasonable space by projecting the allowed space of ($B^{\frac{1}{4}}$, $\tilde{\mu}$, $\Gamma$) to $\Gamma$-$B^{\frac{1}{4}}$ plain and $\tilde{\mu}$-$B^{\frac{1}{4}}$ plain, respectively. We find that $B^{\frac{1}{4}}$ is well constrained to a range of (166.16, 171.06) MeV, differing from the result of (134.1, 141.4) MeV in Ref.~\cite{PhysRevD.97.083015} and $\{$(140, 143) MeV, for $a_4=0.5$; (147, 155) MeV, for $a_4=0.6\}$ in Ref.~\cite{0004-637X-857-1-12}. In addition to that, different value of $B^{\frac{1}{4}}$ can result in different parameter space of ($\tilde{\mu}, \Gamma$). Therefore, we set $B^{\frac{1}{4}}=$167 MeV, 170 MeV, and 171 MeV respectively to study the difference. Then we find that as $B^{\frac{1}{4}}$ increases, the restricted parameter space ($\tilde{\mu}, \Gamma$) is moving to the upper right along the line of $\tilde{\mu}-\Gamma=1$, and becomes larger first and then shrinks. For a detailed study of the constrained hybrid EOS, we choose nine representative parameter sets to calculate their corresponding sound velocities, $M-R$ relation and tidal deformability. As a result, these representative hybrid EOSs are relatively soft but with the maximum mass of HSs well beyond 2 $M_{\odot}$ and radius about 12 km. By a comparison of the phase transition window $\tilde{\mu}-\Gamma\lesssim\mu\lesssim\tilde{\mu}+\Gamma$ and the central baryon chemical potential of 1.17 $M_{\odot}$, 1.36 $M_{\odot}$, 1.59 $M_{\odot}$, and $M_{\rm max}$, we can see that both two member stars of BNS from GW170817 are HSs, and they do not have a quark core but a mixed-phase in center. What's more, the NL3$\omega\rho$ model to construct the pure neutron star has already been excluded by the observation of tidal deformability from GW170817, but this model is still suggested to be effective to describe the hadronic phase in HSs.

As further point it should be noted that we also considered the possibility of an hybrid EOS constructed with the NL3 hadronic model but could not find a parameter set satisfying the five constraints presented in this paper. In addition, the Maxwell construction between hadronic phase and quark phase can be viewed as a limit situation of $\Gamma=0$ and $\tilde{\mu}$ fixed to the intersection of quark EOS and hadronic EOS in $P-\mu$ plane. From Fig.~\ref{Fig:bmugamma}(a), we can see that the parameter space implies $\Gamma\neq0$, thus hybrid EOSs constructed with NL3$\omega\rho$ model and 2+1 flavors NJL model by this approach should be excluded.

In a word, calculations of the hybrid EOS are still model-dependent, but two prospects are hopeful in the future: on one hand, a better constrained tidal deformability from the future observation of GW will help the further reduction of the parameter space; on the other hand, the determination of hadron-quark transition point $\mu_{\rm deconfinement}$ and the EOS from the first principle of QCD in future are expected to give a definitive answer.
\acknowledgments
This work is supported in part by the National Natural Science Foundation of China (under Grants No. 11475085, No. 11535005, No. 11690030, No. 11473012, and No. 11873030), the Fundamental Research Funds for the Central Universities (under Grant No. 020414380074), the National Basic Research Program of China (``973'' Program, Grant No. 2014CB845800), the Strategic Priority Research Program of the Chinese Academy of Sciences
``Multi-waveband Gravitational Wave Universe'' (Grant No. XDB23040000), the National Major state Basic Research and Development of China (Grant No. 2016YFE0129300), the National Post-doctoral Program for Innovative Talents (Grant No. BX201700115), and by the China Postdoctoral Science Foundation funded project (Grant No. 2017M620199).
\section{Appendix: DERIVATION OF QUARK CONDENSATE}\label{five}
In QCD, quark condensate is defined in the Minkowski space. However, it is noteworthy that nonperturbative theories are always proposed and calculated in the Euclidean space, such as lattice QCD (LQCD), because Euclidean QCD action at zero chemical potential defines a probability measure where various numerical simulation algorithms are available. What's more, calculating in the Euclidean space is not only for pragmatic: Euclidean lattice field theory is considered as a primary candidate currently for rigorous definition of the interacting quantum field theory since it makes the definition of generating functional via a proper limiting procedure possible~\cite{Roberts2000S1}. Thus we will take a Wick rotation to translate calculations from the Minkowski space to the Euclidean space. In addition, we also introduce PTR because the Lagrangian of NJL model cannot be renormalized. The PTR is defined as,
\begin{eqnarray}
  \frac{1}{A^n} &=& \frac{1}{(n-1)!}\int_{0}^{\infty}{\rm d}\tau\tau^{n-1}e^{-\tau A}\nonumber \\
   & &\xrightarrow{\rm UV cutoff} \frac{1}{(n-1)!}\int_{\tau_{\rm UV}}^{\infty}{\rm d}\tau\tau^{n-1}e^{-\tau A}.\,\,\label{regularization}
\end{eqnarray}
With the two operations above, the quark condensate defined in Eq.~(\ref{qcondensate}) at zero temperature and chemical potential becomes
\begin{eqnarray}
   \langle\bar{\psi}\psi\rangle_{\rm i} &=& -N_{\rm c}\int_{-\infty}^{+\infty}\frac{{\rm d}^4p^{\rm E}}{(2\pi)^4}\frac{4iM_i}{(p^{\rm E})^{2}+M_i^2}\nonumber\\
   &=& -\frac{N_{\rm c}}{(2\pi)^4}\int_{-\infty}^{+\infty}\int_{-\infty}^{+\infty}{\rm d}^3\overrightarrow{p}{\rm d}p_4\frac{4M_i}{p_4^2+\overrightarrow{p}^2+M_i^2}\nonumber \\
  &=& -\frac{3M_i}{\pi^2}\int_{0}^{+\infty}{\rm d}p\frac{p^2}{\sqrt{p^2+M_i^2}}\nonumber \\
   &=& -\frac{3M_i}{\pi^{\frac{2}{5}}}\int_{\tau_{\rm UV}}^{\infty}\int_{0}^{+\infty}{\rm d}\tau {\rm d}p\tau ^{-\frac{1}{2}}p^2e^{-\tau (M_i^2+p^2)}\nonumber \\
   &=& -\frac{3M_i}{4\pi^2}\int_{\tau_{\rm UV}}^{\infty}{\rm d}\tau \frac{e^{-\tau M_i^2}}{\tau^2},\,\,\label{regofqcondensate}
\end{eqnarray}
here the superscript E denotes the Euclidean space.

On account of the temperature of NSs which can be approximated to zero compared with the chemical potential, we have to generalize our calculation to zero temperature and finite chemical potential. In the Euclidean space, it is equivalent to perform a transformation~\cite{PhysRevC.71.015205} of
\begin{equation}\label{muinpfour}
  p_4\rightarrow p_4+i\mu .
\end{equation}
And then we can derive the quark condensate in the following,
\begin{widetext}
\begin{eqnarray}
  \langle\bar{\psi}\psi\rangle_{\rm i} &=& -N_{\rm c}\int_{-\infty}^{+\infty}\frac{{\rm d}^4p}{(2\pi)^4}\frac{4M_{\rm i}}{(p_4+i\mu)^2+M_{\rm i}^2+\overrightarrow{p}^2}\nonumber \\
   &=& -\frac{3M_{\rm i}}{\pi^3}\int_{0}^{+\infty}{\rm d}p\int_{-\infty}^{+\infty}{\rm d}p_4\frac{p^2}{(p_4+i\mu)^2+M_{\rm i}^2+p^2}\nonumber\\
   &=& \left\{
  \begin{array}{lcl}
\displaystyle{-\frac{3M_{\rm i}}{\pi^2}\int_{\sqrt{\mu^2-M_{\rm i}^2}}^{+\infty}{\rm d}p\textstyle{\frac{\left[1-{\rm Erf}(\sqrt{M_{\rm i}^2+p^2}\sqrt{\tau_{\rm UV}})\right]p^2}{\sqrt{M_{\rm i}^2+p^2}}}},\,\,\,\,   M_{\rm i}<\mu\,\,\,\label{mutoqc}\\
\displaystyle{\frac{3M_{\rm i}}{4\pi^2}\left[\textstyle{-M_{\rm i}^2{\rm Ei}(-M_{\rm i}^2\tau_{\rm UV})-\frac{e^{-M_{\rm i}^2\tau_{\rm UV}}}{\tau_{\rm UV}}}\right]},\,\,\,\,\,\,\,\quad\quad M_{\rm i}>\mu
  \end{array}\right.\qquad\qquad
\end{eqnarray}
\end{widetext}
where Ei(x)$=-\int_{-x}^{+\infty}{\rm d}y\frac{e^{-y}}{t}$ is an Exponential Integral function and Erf(x)$=\frac{2}{\sqrt{\pi}}\int_{0}^{x}e^{-\eta^2}{\rm d}\eta$ is the error function. We can see that the quark condensate depends on its constituent mass and chemical potential. Specifically, for $\mu<M_{\rm i}$, the quark condensate is independent of chemical potential, just like the result in Ref.~\cite{PhysRevD.58.096007}.
\bibliography{reference}
\end{document}